\documentclass[11pt,a4paper]{article}

\usepackage[margin=1in]{geometry}
\usepackage[T1]{fontenc}
\usepackage[utf8]{inputenc}
\usepackage[main=english]{babel}
\usepackage{lmodern}
\usepackage[nopatch=item]{microtype}

\usepackage{amsmath,amsfonts,amssymb}
\usepackage{bm}
\usepackage{algorithm}
\usepackage{algorithmic}
\usepackage{array}
\usepackage{booktabs}
\usepackage{multirow}

\usepackage{graphicx}
\usepackage[font=small,labelfont=bf]{caption}
\usepackage[font=footnotesize]{subcaption}
\usepackage{stfloats}
\usepackage{textcomp}

\usepackage{url}
\usepackage{verbatim}
\usepackage{xcolor}
\usepackage{hyperref}
\usepackage{authblk}
\usepackage{titlesec}
\usepackage{abstract}

\IfFileExists{orcidlink.sty}{%
  \usepackage{orcidlink}%
}{%
  \providecommand{\orcidlink}[1]{\href{https://orcid.org/#1}{\small\textsuperscript{ORCID:#1}}}%
}

\graphicspath{{./figs/}{./}}
\hypersetup{colorlinks=true, linkcolor=blue!60!black,
            citecolor=blue!60!black, urlcolor=blue!60!black}
\newcommand{\doi}[1]{\url{https://doi.org/#1}}

\providecommand{\keywords}[1]{\noindent\textbf{Keywords:} #1\par\medskip}

\title{\bfseries Quantum Machine Learning for Cyber-Physical Anomaly
Detection in Unmanned Aerial Vehicles: A Leakage-Free Evaluation with
Proxy-Audited Feature Sets}

\author[1,2]{Carlos A. Dur\'{a}n Paredes\,\orcidlink{0009-0008-3243-7684}}
\author[3]{Javier E. Le\'{o}n Calder\'{o}n\,\orcidlink{0009-0001-5799-5454}}
\author[4]{Nicol\'{a}s S\'{a}nchez Perea\,\orcidlink{0009-0004-5573-4674}}
\author[5]{German Dar\'{i}o D\'{i}az\,\orcidlink{0009-0004-0032-0618}}
\author[1,2]{Camilo Segura\,\orcidlink{0000-0003-3398-5670}}

\affil[1]{Corporation for Aerospace Initiatives, Research and Innovation
(CASIRI), Popay\'{a}n, Colombia}
\affil[2]{\texttt{caduranpd@gmail.com}, \texttt{camilosegura6@gmail.com}}
\affil[3]{Department of Electronics Engineering, Universidad Nacional de
Colombia, Manizales, Colombia. \texttt{javleonca@unal.edu.co}}
\affil[4]{Department of Electronics Engineering, Universidad del Cauca,
Popay\'{a}n, Colombia. \texttt{nicolassp@unicauca.edu.co}}
\affil[5]{Department of Physics, Universidad del Cauca, Popay\'{a}n,
Colombia. \texttt{germandiaz@unicauca.edu.co}}

\date{\today}

\begin{document}

\maketitle

\begin{abstract}
\noindent Unmanned aerial vehicles (UAVs) are cyber-physical systems whose
attack surface spans networked avionics and on-board sensor fusion: a
compromised GPS or battery module can mimic a benign mission segment and
evade naive anomaly detectors. We present a leakage-free evaluation of
quantum machine learning for UAV anomaly detection on the multi-sensor
TLM:UAV benchmark~\cite{tlmuav2023}. Three contributions support the
study. \emph{(i)}~A group-aware temporal protocol (B2) partitions the
dataset into ten contiguous \texttt{TimeUS} blocks and evaluates over ten
seeds, eliminating the inflation produced by random stratified splits
that mix neighbouring samples. \emph{(ii)}~A three-mode feature audit
(\emph{full}/\emph{loose}/\emph{strict}) quantifies how much accuracy
stems from instantaneous physical signals versus contextual proxies
(cumulative energy, battery state, GPS trajectory). \emph{(iii)}~A hybrid
XGBoost\,+\,Data Re-uploading (DRU) classifier is benchmarked against
five paired non-linear controls (raw, PCA, polynomial-2, random-RBF, and
an untrained DRU map) under identical budgets. The standalone DRU does
not consistently match the strongest classical baseline across seeds;
however, the trained-DRU hybrid is the only model whose mean F1 macro
shifts upward from full to strict ($+0.05$), a directional signal that
the per-seed standard deviations (Table~\ref{tab:headline}) prevent from
being interpreted as a statistically established difference. The
\emph{trained-DRU hybrid} also records the lowest mean false-alarm rate
under proxy-free evaluation, subject to the inter-seed variance reported
in Table~\ref{tab:headline}. We frame this as an incremental,
reproducible \emph{quantum-enhanced hybrid benefit}, and provide an open
Qiskit\,2.x implementation as a benchmark for cybersecurity analytics in
NISQ-era aerospace systems.
\end{abstract}


\section{Introduction}
\label{sec:intro}
UAVs sit at the intersection of two attack surfaces: a \emph{cyber} layer
of networked telemetry and command-and-control links, and a
\emph{physical} layer of sensors and actuators whose readings can be
poisoned, spoofed, or degraded by an
adversary~\cite{enisa2021aviation,kalinin2023security,sedjelmaci2017intrusion}.
A spoofed GPS lock or a tampered IMU can drive an on-board controller to
misclassify a hostile manoeuvre as nominal, making anomaly detection a
first-line cybersecurity control for autonomous aerial systems. The same
telemetry exposes two methodological hazards that are under-reported:
\emph{temporal data leakage} when evaluation shuffles samples across
neighbouring instants, and \emph{contextual proxy features} (cumulative
battery energy, integrated altitude, GPS position) that correlate
trivially with the temporal segment in which a fault was injected.
Either hazard inflates reported scores and yields detectors that fail in
flight.

Quantum machine learning (QML) has been proposed as a complementary tool
for cybersecurity~\cite{biamonte2017quantum,kalinin2023security}, with
quantum feature maps projecting data into Hilbert spaces where non-linear
decision boundaries become more
separable~\cite{havlicek2019supervised,schuld2019quantum}. The data
re-uploading (DRU) paradigm~\cite{perezsalinas2020data} permits compact
variational circuits with few parameters, attractive for embedded
avionics. Yet existing QML-for-cybersecurity studies are dominated by
network-intrusion benchmarks and rarely validate on cyber-physical
telemetry with the protocol rigour expected from aerospace
evaluation~\cite{choudhary2020intrusion}.

This article addresses the gap with a single guiding question:
\textit{under a leakage-free protocol that explicitly audits contextual
proxies, does a quantum or quantum-augmented hybrid classifier offer a
measurable, defensible benefit over deterministic and random non-linear
baselines for UAV anomaly detection?}

\paragraph{Contributions.}
(i)~A \emph{group-aware temporal protocol} (B2) for the multi-sensor
TLM:UAV benchmark with $K\!=\!10$ contiguous \texttt{TimeUS} blocks and
10 seeds.
(ii)~A \emph{three-mode proxy audit} (\emph{full}/\emph{loose}/\emph{strict})
exposing each model's reliance on contextual signals.
(iii)~A \emph{paired-control hybrid analysis} comparing the trained DRU
augmentation against raw, PCA, polynomial-2, random-RBF and an
untrained-DRU map.
(iv)~An \emph{open Qiskit\,2.x} implementation of the DRU classifier as a
\texttt{scikit-learn} estimator~\cite{druqiskit2025}.

\subsection*{Related Work}
Classical UAV intrusion detection has matured around supervised ensembles
and recurrent networks trained on flight-log telemetry, with surveys
documenting both the methodological diversity and the recurring
evaluation pitfalls~\cite{choudhary2020intrusion,sedjelmaci2017intrusion}.
On the quantum side, variational classifiers have been benchmarked on
network-intrusion corpora such as NSL-KDD and
CICIDS~\cite{kalinin2023security,abbas2021power}, and quantum-kernel
methods have shown class-separation gains in feature Hilbert
spaces~\cite{havlicek2019supervised,schuld2019quantum}. Variational
quantum models on near-term hardware operate firmly within the NISQ
regime~\cite{preskill2018nisq,cerezo2021variational}, where unavoidable
noise constrains the achievable accuracy and motivates hybrid
quantum--classical architectures. Two gaps motivate the present study.
First, cyber-physical UAV telemetry is rarely evaluated under
group-aware temporal protocols, so reported scores are likely optimistic
relative to in-flight deployment. Second, hybrid quantum augmentations
are seldom compared against \emph{paired} non-linear controls
(deterministic and random) of equivalent expressive budget, leaving the
contribution of the variational circuit confounded with the contribution
of generic feature expansion~\cite{abbas2021power,schuld2021supervised}.
We close both gaps within a single reproducible pipeline.

\section{Methodology}
\label{sec:method}
\begin{figure}[!t]
  \centering
  \includegraphics[width=0.85\linewidth]{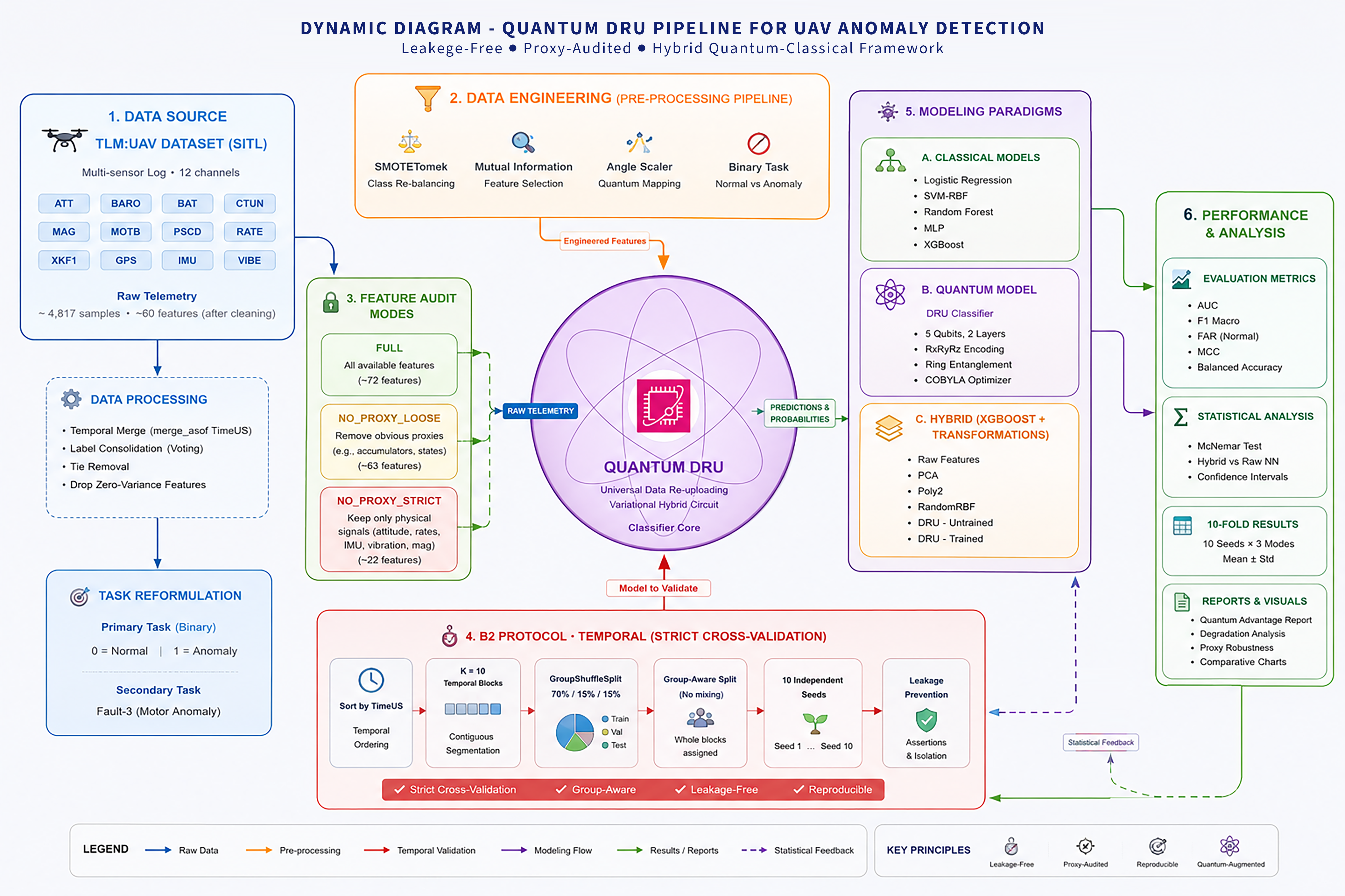}
  \caption{Hub-and-spoke architecture of the experimental pipeline. The
    central hub implements the leakage-free preprocessing core: temporal
    ordering by \texttt{TimeUS}, $K$-block group-aware split, train-only
    scaling and balancing (RobustScaler $\rightarrow$ SMOTETomek), and
    mutual-information feature ranking under the three audit modes
    (\emph{full}/\emph{loose}/\emph{strict}). From this hub, three model
    spokes are evaluated under identical seeds, splits, and budgets:
    (i)~classical baselines (Logistic Regression, SVM-RBF, MLP, Random
    Forest, XGBoost), (ii)~the standalone quantum DRU classifier
    (5~qubits, 2 layers, $R_xR_yR_z$ encoding with ring entanglement),
    and (iii)~the hybrid family that augments XGBoost with raw, PCA,
    polynomial-2, random-RBF, untrained-DRU, or trained-DRU features.
    The architecture is designed so that any observed difference between
    paradigms reflects the representation, not the evaluation harness.}
  \label{fig:hub}
\end{figure}

Fig.~\ref{fig:hub} summarises the experimental pipeline as a hub-and-spoke
architecture: a leakage-free preprocessing core feeds the classical,
quantum, and hybrid spokes under identical evaluation conditions.

\subsection{Dataset and Task Reformulation}
TLM:UAV~\cite{tlmuav2023} bundles software-in-the-loop telemetry from
twelve sensor subsystems (ATT, BARO, BAT, CTUN, MAG, MOTB, PSCD, RATE,
XKF1, GPS, IMU, VIBE) with per-sample labels for four anomaly types
(1~GPS; 2~accelerometer; 3~engine; 4~RC) plus nominal operation
(0~normal). The original Time Line Modeling (TLM) methodology anchors
fault intervals from the simulated flight timeline, stretches abnormal
windows to mitigate class imbalance, and explicitly removes time-related
or non-universal features before model training~\cite{tlmuav2023}. A
later multi-sensor study on the same UAV anomaly-detection setting
further emphasises that heterogeneous sensor alignment is itself a
modelling step, not a neutral preprocessing detail~\cite{deng2024ifhmnn}.

\paragraph{Fusion-table integrity audit.}
The Kaggle release also includes a convenience table,
\path{Fusion_Data.csv}. We do not use it as the primary experimental
source. A file-level integrity audit found two exact duplicate pairs in
that table, \texttt{ErrRP}$=$\texttt{ErrYaw} and
\texttt{MagY}$=$\texttt{MagZ}. When the same pairs were checked in the
raw sensor files, they were not exact duplicates
(\texttt{ErrRP}/\texttt{ErrYaw} same-ratio $\approx 0.062$ and
\texttt{MagY}/\texttt{MagZ} same-ratio $=0$). This pattern is more
consistent with a fusion/export artefact than with genuine physical
redundancy in the raw logs. The same audit also found high sensitivity
to row-wise random splitting, confirming that the dataset must be
evaluated through temporal or group-aware partitions rather than by
shuffling neighbouring samples.

Consequently, our pipeline reconstructs the working table from raw
sensor files instead of relying on \path{Fusion_Data.csv}. Sources are
aligned on \texttt{TimeUS} (\texttt{merge\_asof} for high-rate streams),
naming collisions are resolved by explicit sensor-aware renames,
per-sensor labels are voted into a per-row label, and tied rows are
discarded. This choice makes the data-generation assumptions visible and
keeps the evaluation consistent with the TLM principle that time markers
and mission-specific coordinates should not be treated as direct
predictors. The resulting table holds 4\,817 samples and 72 numeric
features after dropping zero-variance columns.

\begin{figure}[!t]
  \centering
  \includegraphics[width=0.85\linewidth]{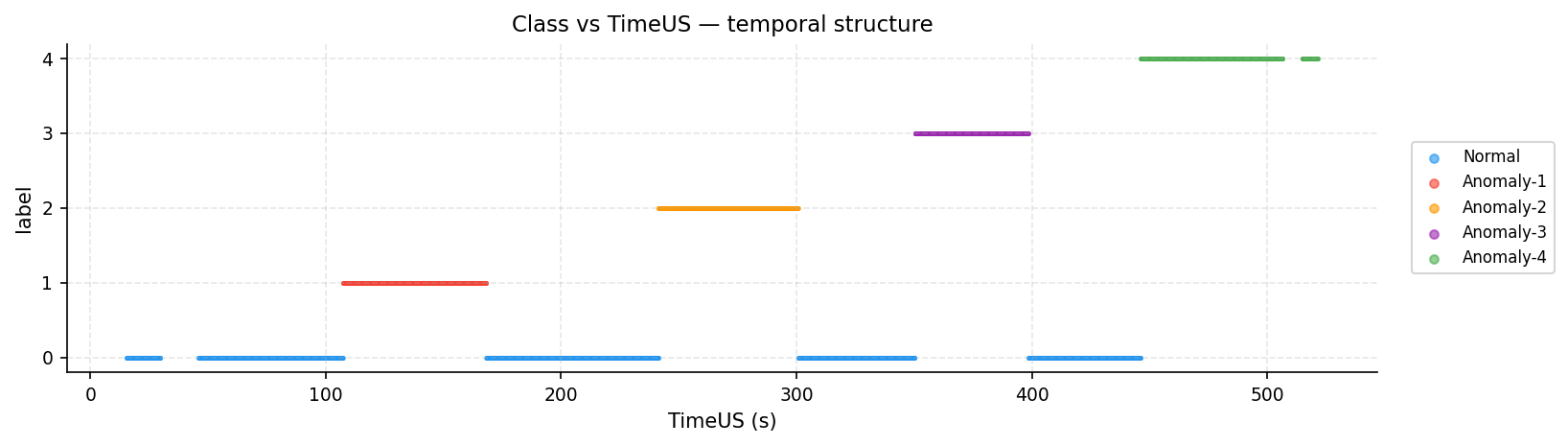}
  \caption{TLM:UAV anomaly classes along \texttt{TimeUS}. Each anomaly
    type is confined to one or two segments, with only three gap-defined
    episodes.}
  \label{fig:timeline}
\end{figure}

A temporal diagnostic (Fig.~\ref{fig:timeline}) reveals only \emph{three
disjoint episodes} defined by gaps in \texttt{TimeUS}, each carrying a
different subset of anomaly classes. Splitting three episodes across
train/validation/test under group-aware sampling leaves at least one
split with a missing class, so the multiclass task is structurally
infeasible. We therefore reformulate the primary task as \emph{binary
anomaly detection} ($y\!=\!1$ if any fault is present, $0$ otherwise) and
retain \emph{Fault-3} (Normal vs.\ motor anomaly) as a secondary
analysis.

\subsection{B2 Group-Aware Protocol and Per-Seed Pipeline}
\label{sec:b2}
For every seed $s\!\in\!\{0,\dots,9\}$ we (i)~order by \texttt{TimeUS},
(ii)~cut into $K\!=\!10$ contiguous blocks, and (iii)~apply a two-stage
\texttt{GroupShuffleSplit} that assigns whole blocks (never rows) to a
70/15/15 train/validation/test partition, enforced by assertion. Across
seeds the prior shift between splits varies from 0.13 to 0.72; reporting
mean\,$\pm$\,std stress-tests each model under deployment-time prior
shift. Each $(s,\text{mode})$ pair runs an isolated pipeline: feature
subset~$\rightarrow$~RobustScaler (fit on train
only)~$\rightarrow$~SMOTETomek on the training
fold~$\rightarrow$~Mutual-Information ranking on the balanced training
set~$\rightarrow$~top-5 selection~$\rightarrow$~MinMax angle scaler to
$[-\pi,\pi]$. No transformer is fit globally; no synthetic sample reaches
validation or test.

\subsection{Three Feature-Audit Modes}
\label{sec:proxy}
\textbf{full} retains all 72 non-zero-variance features.
\textbf{loose} drops nine accumulators and state flags
(\texttt{abT, EnrgTot, CurrTot, Res, BatRes, Offset, Rout, POut, YOut};
63 features). \textbf{strict} additionally excludes battery state,
GPS/position-estimator outputs, controller setpoints, altitude/baro and
motor demand; the 22 surviving features are attitude
(\texttt{Roll, Pitch, Yaw}), body-frame rates and gyro-bias estimates,
IMU accelerations and angular rates, magnetometer and vibration
channels. An MI-stability audit confirms the design: in \emph{full} the
top-five features are \emph{all} proxies (\texttt{Offset, abT, CurrTot,
EnrgTot, BatRes}, inclusion rate~1.0); only in \emph{strict} do the
top-MI features become physical (\texttt{Yaw, GX, MagY, GY, GZ}).

\subsection{Models}
\textbf{Classical baselines}: Logistic Regression, SVM-RBF, Random
Forest, MLP, and XGBoost~\cite{chen2016xgboost}, trained on the
post-SMOTETomek balanced fold. \textbf{DRU}: 5 qubits, 2 layers,
$R_xR_yR_z$ encoding, ring entanglement, 30 trainable parameters, COBYLA
optimiser, training budget $|A|\!\leq\!400$ samples per class on a
balanced subset~$A$ disjoint from the hybrid set~$B$ to forbid
information leak between the DRU and its XGBoost
head~\cite{perezsalinas2020data,schuld2019quantum}.
\textbf{Hybrid family} (six variants): XGBoost trained on
$X_q\,\parallel\,T(X_q)$, with $T\in\{$raw, PCA, Poly$^2$, RandomRBF,
DRU-untrained, DRU-trained$\}$. The Quantum Kernel SVM (\texttt{QSVC}
with ZZFeatureMap)~\cite{havlicek2019supervised} is implemented but
disabled by default because each fit exceeds 40\,min on the per-seed
test fold.

\section{Results}
\label{sec:results}

\subsection{Headline Comparison}
\begin{figure}[!t]
  \centering
  \includegraphics[width=0.85\linewidth]{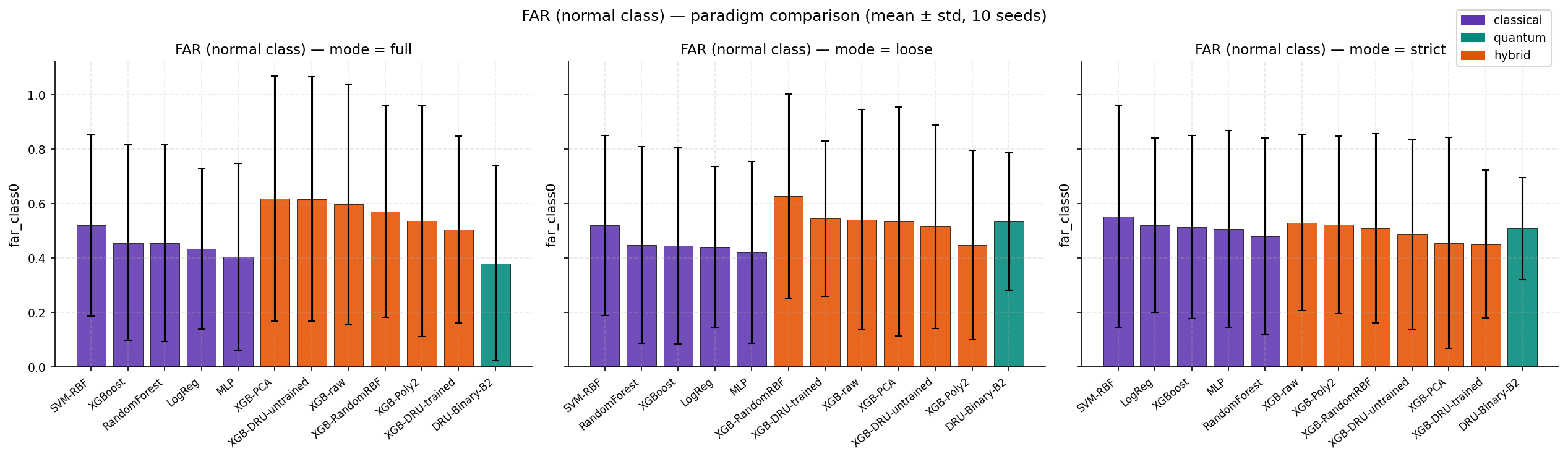}
  \caption{False-alarm rate (FAR) on the normal class across paradigms.
    Under \emph{strict}, the trained-DRU hybrid attains the lowest FAR
    (0.451), a key operational metric for an intrusion-style detector.}
  \label{fig:far}
\end{figure}

Table~\ref{tab:headline} reports mean\,$\pm$\,std across ten seeds for
the \emph{full} and \emph{strict} modes (\emph{loose} omitted for space;
see supplementary material). Figs.~\ref{fig:f1}, \ref{fig:auc}, and
\ref{fig:far} render the same data as cross-paradigm bar charts for the
three operationally relevant metrics.

\begin{figure}[!t]
  \centering
  \includegraphics[width=0.85\linewidth]{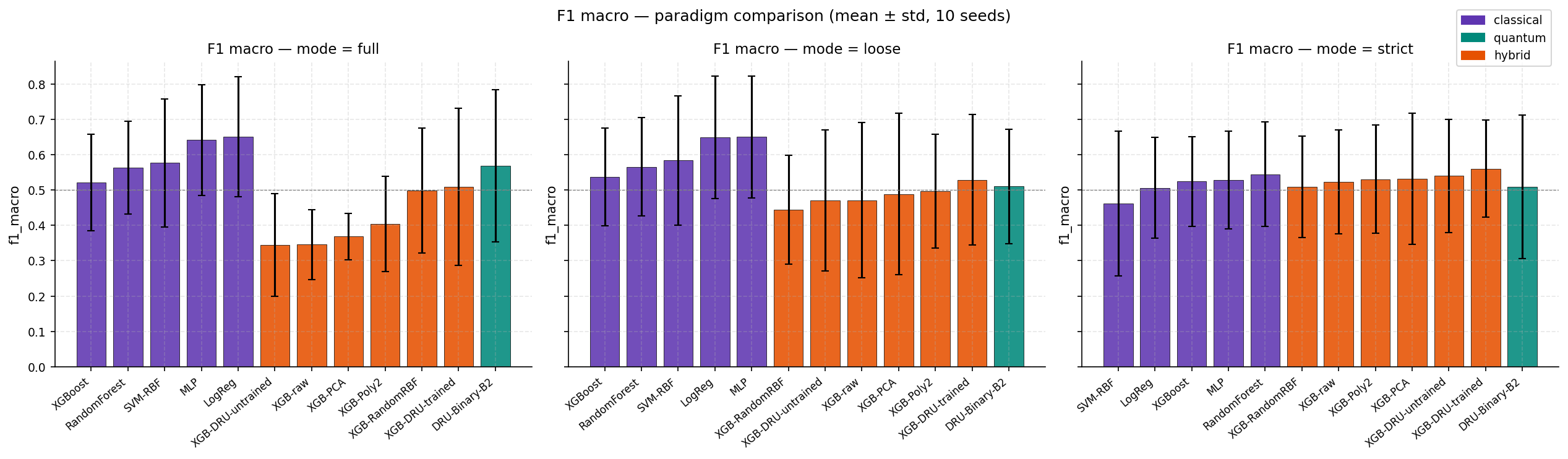}
  \caption{F1 macro across paradigms, per feature mode. Classical models
    carry most of their performance from contextual proxies and degrade
    visibly from \emph{full} to \emph{strict}; the trained-DRU hybrid
    (rightmost) is the only model that \emph{improves} under
    \emph{strict}.}
  \label{fig:f1}
\end{figure}

\begin{figure}[!t]
  \centering
  \includegraphics[width=0.85\linewidth]{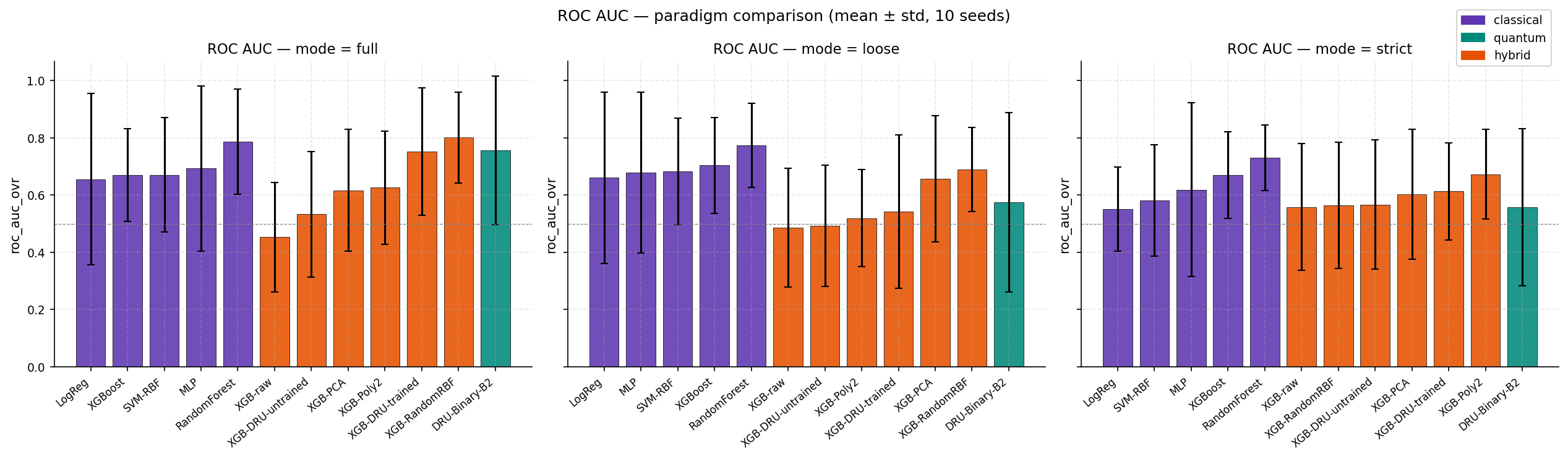}
  \caption{ROC AUC across paradigms. Random Forest is the most
    proxy-robust classical baseline; the standalone DRU is competitive
    in \emph{full} (0.76) but degrades sharply, indicating that its raw
    representation still benefits from contextual signals.}
  \label{fig:auc}
\end{figure}

\begin{table}[!t]
\centering
\caption{TLM:UAV binary task under B2 (mean\,$\pm$\,std, 10 seeds). Best
per column in bold; trained-DRU hybrid highlighted.}
\label{tab:headline}
\setlength{\tabcolsep}{4.5pt}
\renewcommand{\arraystretch}{1.05}
\small

\resizebox{\textwidth}{!}{%
\begin{tabular}{ll cc cc cc}
\toprule
& & \multicolumn{2}{c}{\textbf{F1 macro}}
  & \multicolumn{2}{c}{\textbf{ROC AUC}}
  & \multicolumn{2}{c}{\textbf{FAR (normal)} $\downarrow$} \\
\cmidrule(lr){3-4}\cmidrule(lr){5-6}\cmidrule(lr){7-8}
\textbf{Paradigm} & \textbf{Model} & full & strict & full & strict & full & strict \\
\midrule
Classical & Logistic Reg.  & $\bm{0.650\pm0.17}$ & $0.506\pm0.14$ & $0.655\pm0.30$ & $0.551\pm0.15$ & $0.433\pm0.29$ & $0.520\pm0.32$ \\
Classical & MLP            & $0.641\pm0.16$       & $0.528\pm0.14$ & $0.693\pm0.29$ & $0.619\pm0.31$ & $0.405\pm0.34$ & $0.507\pm0.36$ \\
Classical & Random Forest  & $0.564\pm0.13$       & $0.545\pm0.15$ & $0.787\pm0.18$ & $\bm{0.731\pm0.11}$ & $0.454\pm0.36$ & $0.480\pm0.36$ \\
Classical & SVM-RBF        & $0.577\pm0.18$       & $0.462\pm0.20$ & $0.670\pm0.20$ & $0.581\pm0.19$ & $0.520\pm0.33$ & $0.553\pm0.41$ \\
Classical & XGBoost        & $0.521\pm0.14$       & $0.524\pm0.13$ & $0.669\pm0.16$ & $0.669\pm0.15$ & $0.455\pm0.36$ & $0.513\pm0.34$ \\
\midrule
Quantum   & DRU-Binary     & $0.568\pm0.22$       & $0.510\pm0.20$ & $0.757\pm0.26$ & $0.557\pm0.27$ & $\bm{0.380\pm0.36}$ & $0.509\pm0.19$ \\
\midrule
Hybrid    & XGB-raw            & $0.345\pm0.10$ & $0.523\pm0.15$ & $0.453\pm0.19$ & $0.558\pm0.22$ & $0.597\pm0.44$ & $0.530\pm0.32$ \\
Hybrid    & XGB-PCA            & $0.368\pm0.07$ & $0.532\pm0.19$ & $0.616\pm0.21$ & $0.603\pm0.23$ & $0.619\pm0.45$ & $0.455\pm0.39$ \\
Hybrid    & XGB-Poly$^2$       & $0.404\pm0.14$ & $0.531\pm0.15$ & $0.626\pm0.20$ & $0.673\pm0.16$ & $0.535\pm0.42$ & $0.522\pm0.33$ \\
Hybrid    & XGB-RandomRBF      & $0.498\pm0.18$ & $0.509\pm0.14$ & $\bm{0.801\pm0.16}$ & $0.564\pm0.22$ & $0.570\pm0.39$ & $0.510\pm0.35$ \\
Hybrid    & XGB-DRU-untrained  & $0.345\pm0.14$ & $0.540\pm0.16$ & $0.533\pm0.22$ & $0.566\pm0.23$ & $0.617\pm0.45$ & $0.487\pm0.35$ \\
Hybrid    & \textbf{XGB-DRU-trained} & $0.509\pm0.22$ & $\bm{0.561\pm0.14}$ & $0.751\pm0.22$ & $0.613\pm0.17$ & $0.504\pm0.34$ & $\bm{0.451\pm0.27}$ \\
\bottomrule
\end{tabular}%
}

\vspace{0.4em}
\begin{minipage}{0.97\linewidth}\footnotesize
B2 protocol: $K\!=\!10$ TimeUS blocks, 10 seeds, binary task.
\emph{full}: 72 features; \emph{strict}: 22 physical features
(Sec.~\ref{sec:proxy}).
\end{minipage}
\end{table}

\subsection{Degradation Profile and Hybrid Controls}
\label{sec:results-hybrid}
Most classical models degrade from \emph{full} to \emph{strict}:
$\Delta$F1 ranges from $-0.14$ (Logistic) to $-0.02$ (Random Forest).
The standalone DRU also degrades ($\Delta$F1$=-0.06$,
$\Delta$AUC$=-0.20$): its raw representation is not itself proxy-free,
since the top-MI angles still encode contextual signals. Random Forest
shows the highest mean ROC AUC under \emph{strict} and the smallest
degradation slope among classical models.

\begin{figure}[!t]
  \centering
  \includegraphics[width=0.85\linewidth]{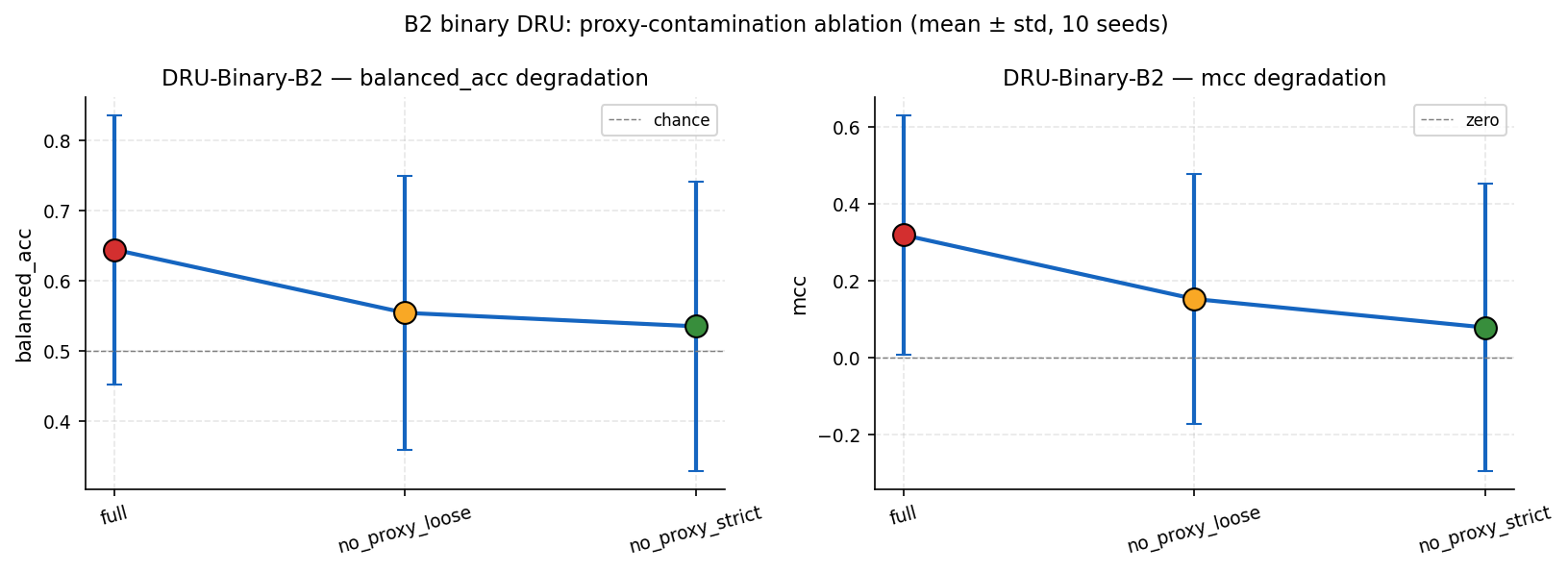}
  \caption{DRU degradation from \emph{full} to \emph{strict}. The slope
    on F1 macro is $-0.06$; on MCC it is $-0.24$. The DRU's MI-selected
    angles still encode part of the contextual proxies, motivating the
    hybrid augmentation.}
  \label{fig:dru_deg}
\end{figure}

Fig.~\ref{fig:dru_deg} plots the DRU degradation slope explicitly: F1
falls from $0.57\!\pm\!0.22$ in \emph{full} to $0.51\!\pm\!0.20$ in
\emph{strict} (slope $-0.06$), while MCC degrades by $0.24$. Raw quantum
representations alone therefore do not survive aggressive proxy removal,
but they retain enough discriminative geometry to be useful as a feature
provider for a downstream tree ensemble.

\begin{figure}[!t]
  \centering
  \includegraphics[width=0.85\linewidth]{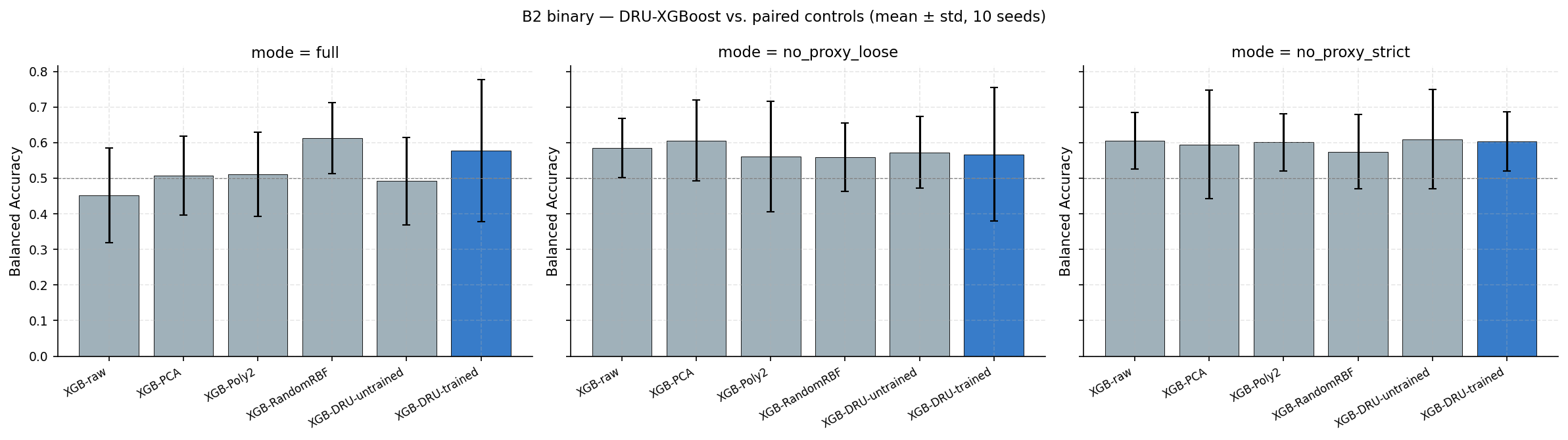}
  \caption{Hybrid XGBoost variants under the three feature modes. The
    trained-DRU variant (last bar of each panel) is the only one whose
    balanced accuracy stays competitive across modes; PCA, Poly$^2$, and
    random-RBF behave similarly to one another, suggesting that the
    trained DRU contributes information not present in deterministic or
    random non-linear expansions.}
  \label{fig:hyb}
\end{figure}

A cybersecurity-relevant finding lies in Fig.~\ref{fig:hyb}. In every
feature mode the \emph{trained}-DRU hybrid shows a higher mean F1 macro
than three of the five paired controls (PCA, Poly$^2$, and random-RBF),
with mean differences in the range 0.02--0.05 and overlapping standard
deviations (Table~\ref{tab:headline}), and is the only model whose F1
\emph{rises} in \emph{strict} ($+0.05$, from 0.509 to 0.561). It also
records the lowest mean FAR among all models in \emph{strict}
($0.451\!\pm\!0.27$); the large standard deviation reflects inter-seed
prior shift variability and prevents a conclusive ranking. FAR remains a
key operational metric for an intrusion-style detector. The
untrained-DRU map does not reproduce this behaviour, so the directional
difference between trained and untrained DRU maps is consistent with a
contribution from the variational parameters, though overlapping
standard deviations prevent a strong causal
attribution~\cite{abbas2021power}.

\subsection{Fault-3 Secondary Task}
Of the four anomaly classes only \emph{Anomaly-Motor} (label~3) is
evaluable in isolation under group-aware splitting because its samples
span multiple temporal blocks. With $K\!=\!20$ blocks and XGBoost,
\emph{full}/\emph{loose} attain F1\,$=$\,0.93 / Bal.Acc\,$=$\,0.96 (proxy
dominated); under \emph{strict} the task collapses to Bal.Acc\,$=$\,0.50,
F1\,$=$\,0.43, mirroring the binary finding. This is itself a
cybersecurity-relevant result: removing context-laden telemetry can
leave a defender unable to distinguish a motor anomaly from a benign
manoeuvre, exposing a single-point-of-failure in the feature pipeline.

\section{Discussion}
\label{sec:discussion}
The picture is more nuanced than the binary `quantum advantage / no
advantage' framing common in QML papers. The standalone DRU does not
consistently match the strongest classical baseline across seeds or
modes, and Random Forest shows the highest mean ROC AUC under
\emph{strict}. However, the trained-DRU hybrid shows higher mean strict
F1 than three of the five paired controls (PCA, Poly$^2$, and
random-RBF) and the lowest mean FAR in that mode. We interpret this as a
directional signal consistent with an incremental quantum-enhanced
hybrid benefit, pending confirmation under a larger or more episodically
diverse dataset and formal significance testing on a metric central to
aerospace operations: the false-alarm rate when the model is forced to
work from physical signal rather than contextual proxies. False alarms
drive operator desensitisation, and proxy-dependent detectors lose
effectiveness the moment a vehicle is flown in a regime not represented
in training~\cite{choudhary2020intrusion}.

Read only in \emph{full}, several classical models reach
F1$\,\approx\,$0.65 and the DRU's ROC AUC of 0.76 superficially looks
like quantum-advantage evidence. The three-mode audit shows that those
classical models drop by 0.10--0.14 F1 once cumulative and state features
are removed, that the DRU drops by 0.06 F1 and 0.20 AUC, and that the
only model whose F1 \emph{improves} is the trained-DRU hybrid. The audit
is the mechanism that separates a detector which has learned a fault
signature from one that has memorised the mission segment in which the
fault was injected: the AI-for-cybersecurity equivalent of learning the
threat versus learning the test harness.

The fusion-table audit is central to this interpretation. Had the
released \path{Fusion_Data.csv} table been used without inspection,
the duplicated feature pairs and row-wise temporal mixing would have made
the benchmark appear substantially cleaner than it is. Reconstructing
the table from raw logs does not merely change implementation details;
it changes the scientific question from ``can a classifier exploit a
convenient fused table?'' to ``can a representation generalise under
audited sensor fusion, proxy removal, and group-aware temporal
evaluation?''

\paragraph{Statistical scope of the reported differences.}
All comparisons are reported as mean $\pm$ standard deviation over ten
seeds, using the same dataset partition for all paradigms to ensure a
fair comparison. Relative uncertainties range from 15\% to 95\% of the
point estimate (Table~\ref{tab:headline}). This variability is
structural: the B2 protocol exposes each seed to a prior shift between
0.13 and 0.72, and the three disjoint temporal episodes of TLM:UAV
produce train/test distributions that differ substantially across seeds.
Classical models with no quantum budget constraint show comparable
variance in FAR (79--84\%), confirming that the source is the evaluation
design rather than any paradigm-specific limitation. Under these
conditions, observed mean differences between models (typically
0.02--0.05 F1) are smaller than or comparable to within-model standard
deviations, and no formal null-hypothesis test has been applied. Results
should therefore be read as directional trends. Future work requires
either additional independent temporal episodes to widen the effective
evaluation window, or a formal paired significance test across seeds.

\paragraph{Limitations.}
(i)~The three-episode footprint of TLM:UAV makes multiclass evaluation
structurally infeasible under group-aware sampling; a dataset extension
with independent episodes per anomaly class is the logical next step.
(ii)~The released \path{Fusion_Data.csv} table contains duplicated
feature pairs not present as exact duplicates in the raw logs, so our
results should be read as applying to the audited raw-reconstructed
table, not to the convenience fusion file. (iii)~All DRU results are
state-vector simulations; deployment on NISQ
hardware~\cite{preskill2018nisq} is expected to introduce additional
accuracy degradation from gate noise, readout errors, and decoherence
in variational
circuits~\cite{cerezo2021variational}, partially absorbed by the hybrid
head; quantifying the simulator-vs-hardware gap on TLM:UAV is left for
future work. Quantum classifiers may additionally be vulnerable to
adversarial perturbations, an orthogonal robustness concern documented
on superconducting qubits~\cite{ren2022quantum}. (iv)~The kernel-SVM
quantum baseline is implemented but disabled by default for runtime
reasons; sub-sampled or trainable quantum
kernels~\cite{schuld2021supervised} are needed for a fair 10-seed
comparison.

\section{Conclusion}
\label{sec:conclusion}
This work asked whether a quantum or quantum-augmented hybrid classifier
offers a measurable, defensible benefit over deterministic and random
non-linear baselines for UAV anomaly detection, once two pervasive
evaluation hazards---temporal leakage and contextual proxy
features---are explicitly controlled. The answer is qualified rather
than categorical, and the qualification is the contribution.

Under the leakage-free, proxy-audited B2 protocol, the standalone DRU
classifier does not consistently match the strongest classical baseline
across seeds, and Random Forest remains the most proxy-robust classical
model under \emph{strict} evaluation. The trained-DRU hybrid is
nonetheless distinguished on two operationally meaningful axes: it is the
only model whose F1 macro \emph{increases} from \emph{full} to
\emph{strict} ($+0.05$, from 0.509 to 0.561), and it attains the lowest
mean false-alarm rate under proxy-free evaluation
($0.451\!\pm\!0.27$). It also exceeds the mean performance of three of
the five paired controls (PCA, Poly$^2$, and random-RBF) and of the
untrained-DRU map across the metrics examined. Because these differences
(typically 0.02--0.05 F1) fall within the inter-seed standard deviation,
we report them as a directionally consistent, reproducible
\emph{quantum-enhanced hybrid benefit} rather than a statistically
established quantum advantage. The contrast between the trained and
untrained DRU maps is the cleanest internal control supporting a genuine
contribution from the variational parameters, though it too is bounded by
inter-seed variance.

The principal transferable result is not a single score but an evaluation
template. The B2 group-aware temporal protocol, the three-mode feature
audit, and the paired-control hybrid comparison jointly separate a
detector that has learned a fault signature from one that has memorised
the mission segment in which a fault was injected. The fusion-table
integrity audit reframes the scientific question from ``can a classifier
exploit a convenient fused table?'' to ``can a representation generalise
under audited sensor fusion, proxy removal, and group-aware temporal
evaluation?''---a distinction that materially changes which conclusions
about quantum benefit are admissible. We argue this template should
precede any quantum-advantage claim on cyber-physical telemetry, where
random stratified splits and accumulator features can inflate reported
scores by 0.10--0.14 F1.

For an intrusion-style detector, the false-alarm rate under proxy-free
conditions is the metric that governs operational trust: persistent false
alarms drive operator desensitisation, and proxy-dependent detectors lose
effectiveness the instant a vehicle is flown in a regime absent from
training~\cite{choudhary2020intrusion}. The Fault-3 collapse under
\emph{strict} evaluation (Bal.Acc $0.96\!\rightarrow\!0.50$) is a concrete
warning: a feature pipeline that silently depends on context-laden
telemetry constitutes a single point of failure that an adversary aware
of the mission profile could exploit.

Three directions follow directly. First, a dataset extension with
independent temporal episodes per anomaly class would restore the
multiclass task and widen the effective evaluation window, enabling formal
paired significance testing across seeds. Second, the simulator-to-NISQ
gap must be quantified on this benchmark under realistic noise
models~\cite{cerezo2021variational,preskill2018nisq}, together with the
adversarial-robustness dimension that is orthogonal to
accuracy~\cite{ren2022quantum}. Third, sub-sampled or trainable quantum
kernels~\cite{schuld2021supervised} would make a fair 10-seed kernel-SVM
comparison tractable. Until then, the present pipeline stands as an open,
reproducible reference point for cybersecurity analytics in NISQ-era
aerospace systems.

\section*{Code, Data, and Reproducibility Resources}
The implementation is split across two openly accessible companion
repositories. The Data Re-uploading (DRU) classifier is released as a
standalone, pip-installable, \texttt{scikit-learn}-compatible Python
package built on Qiskit\,2.x V2 primitives, with continuous-integration
testing, a published \texttt{CITATION.cff}, and a dual
MIT~/~CC\,BY\,4.0 license covering code and documentation
respectively~\cite{druqiskit2025}.\footnote{\url{https://github.com/Carlosandp/qiskit-data-reuploading}}
The end-to-end experimental pipeline that integrates this classifier into
the leakage-free B2 protocol is maintained in a separate research
repository~\cite{tlmuavrepo2026},\footnote{\url{https://github.com/Carlosandp/TLM-UAV-Quantum-Anomaly-Detection}}
released under CC\,BY-NC-SA\,4.0. It contains the end-to-end notebook
(\path{notebooks/TLM_DRU_FINAL.ipynb}), the aggregated per-seed results
(\path{results/summary/final_comparison_aggregated.csv}), the proxy-audit
report, the publication figures reproduced in this article, and
\texttt{methodology} and \texttt{reproducibility} documents specifying
seeds, runtime expectations, and hardware notes. The DRU package is
consumed by the pipeline as an external dependency pinned to a specific
upstream commit, so the two artefacts version independently while
remaining exactly reproducible together.

To respect the licensing of the source benchmark, the full
\path{Fusion_Data.csv} table is \emph{not} redistributed: the data
directory provides the dataset provenance, the expected schema, a
SHA-256 checksum for integrity verification, and a 100-row sample for
smoke-testing the pipeline without the complete dataset. The original
TLM:UAV data must be obtained from its primary
source~\cite{tlmuav2023}. This separation keeps every code path
reproducible while leaving redistribution rights with the dataset
authors.

\section*{Acknowledgments}
The authors thank the \emph{Corporation for Aerospace Initiatives,
Research and Innovation (CASIRI)} for providing the workspace,
computational resources, and institutional support that made the
development of this research possible.

\bibliographystyle{IEEEtran}  
\bibliography{refs}

@misc{enisa2021aviation,
  author       = {{ENISA}},
  title        = {{ENISA Threat Landscape 2021}},
  howpublished = {European Union Agency for Cybersecurity},
  year         = {2021},
  month        = oct,
  note         = {Available: \url{https://www.enisa.europa.eu/publications/enisa-threat-landscape-2021}}
}

@article{biamonte2017quantum,
  author  = {Biamonte, Jacob and Wittek, Peter and Pancotti, Nicola and Rebentrost, Patrick and Wiebe, Nathan and Lloyd, Seth},
  title   = {Quantum machine learning},
  journal = {Nature},
  volume  = {549},
  number  = {7671},
  pages   = {195--202},
  year    = {2017},
  doi     = {10.1038/nature23474}
}

@article{havlicek2019supervised,
  author  = {Havl{\'i}{\v c}ek, Vojt{\v e}ch and C{\'o}rcoles, Antonio D. and Temme, Kristan and Harrow, Aram W. and Kandala, Abhinav and Chow, Jerry M. and Gambetta, Jay M.},
  title   = {Supervised learning with quantum-enhanced feature spaces},
  journal = {Nature},
  volume  = {567},
  number  = {7747},
  pages   = {209--212},
  year    = {2019},
  doi     = {10.1038/s41586-019-0980-2}
}

@article{perezsalinas2020data,
  author  = {P{\'e}rez-Salinas, Adri{\'a}n and Cervera-Lierta, Alba and Gil-Fuster, Elies and Latorre, Jos{\'e} I.},
  title   = {Data re-uploading for a universal quantum classifier},
  journal = {Quantum},
  volume  = {4},
  pages   = {226},
  year    = {2020},
  doi     = {10.22331/q-2020-02-06-226}
}

@article{kalinin2023security,
  author  = {Kalinin, Maxim and Krundyshev, Vasiliy},
  title   = {Security intrusion detection using quantum machine learning techniques},
  journal = {Journal of Computer Virology and Hacking Techniques},
  volume  = {19},
  pages   = {125--136},
  year    = {2023},
  doi     = {10.1007/s11416-022-00435-0}
}

@article{tlmuav2023,
  author  = {Yang, Tao and Lu, Yu and Deng, Hongli and Chen, Jiangchuan and Tang, Xiaomei},
  title   = {Acquisition and processing of {UAV} fault data based on time line modeling method},
  journal = {Applied Sciences},
  volume  = {13},
  number  = {7},
  pages   = {4301},
  year    = {2023},
  doi     = {10.3390/app13074301}
}

@article{schuld2019quantum,
  author  = {Schuld, Maria and Killoran, Nathan},
  title   = {Quantum machine learning in feature {H}ilbert spaces},
  journal = {Physical Review Letters},
  volume  = {122},
  number  = {4},
  pages   = {040504},
  year    = {2019},
  doi     = {10.1103/PhysRevLett.122.040504}
}

@book{schuld2021supervised,
  author    = {Schuld, Maria and Petruccione, Francesco},
  title     = {Machine Learning with Quantum Computers},
  edition   = {2nd},
  publisher = {Springer},
  year      = {2021},
  doi       = {10.1007/978-3-030-83098-4}
}

@article{ren2022quantum,
  author  = {Ren, Wenhui and Li, Weikang and Xu, Shibo and Wang, Ke and Jiang, Wenjie and Jin, Feitong and Zhu, Xuhao and Chen, Jiachen and Song, Zixuan and Zhang, Pengfei and Dong, Hang and Zhang, Xu and Deng, Jinfeng and Gao, Yu and Zhang, Chuanyu and Wu, Yaozu and Zhang, Bing and Guo, Qiujiang and Li, Hekang and Wang, Zhen and Biamonte, Jacob and Song, Chao and Deng, Dong-Ling and Wang, H.},
  title   = {Experimental quantum adversarial learning with programmable superconducting qubits},
  journal = {Nature Computational Science},
  volume  = {2},
  pages   = {711--717},
  year    = {2022},
  doi     = {10.1038/s43588-022-00351-9}
}

@article{abbas2021power,
  author  = {Abbas, Amira and Sutter, David and Zoufal, Christa and Lucchi, Aurelien and Figalli, Alessio and Woerner, Stefan},
  title   = {The power of quantum neural networks},
  journal = {Nature Computational Science},
  volume  = {1},
  number  = {6},
  pages   = {403--409},
  year    = {2021},
  doi     = {10.1038/s43588-021-00084-1}
}

@inproceedings{chen2016xgboost,
  author    = {Chen, Tianqi and Guestrin, Carlos},
  title     = {{XGBoost}: A scalable tree boosting system},
  booktitle = {Proceedings of the 22nd ACM SIGKDD International Conference on Knowledge Discovery and Data Mining},
  pages     = {785--794},
  year      = {2016},
  publisher = {ACM},
  doi       = {10.1145/2939672.2939785}
}

@article{sedjelmaci2017intrusion,
  author  = {Sedjelmaci, Hichem and Senouci, Sidi Mohammed and Ansari, Nirwan},
  title   = {Intrusion detection and ejection framework against lethal attacks in {UAV}-aided networks: A {B}ayesian game-theoretic methodology},
  journal = {IEEE Transactions on Intelligent Transportation Systems},
  volume  = {18},
  number  = {5},
  pages   = {1143--1153},
  year    = {2017},
  doi     = {10.1109/TITS.2016.2600370}
}

@inproceedings{choudhary2020intrusion,
  author    = {Choudhary, Gaurav and Sharma, Vishal and You, Ilsun and Yim, Kangbin and Chen, Ing-Ray and Cho, Jin-Hee},
  title     = {Intrusion detection systems for networked unmanned aerial vehicles: A survey},
  booktitle = {Proceedings of the 14th International Wireless Communications \& Mobile Computing Conference (IWCMC)},
  pages     = {560--565},
  year      = {2018},
  publisher = {IEEE},
  doi       = {10.1109/IWCMC.2018.8450305}
}

@misc{druqiskit2025,
  author       = {Dur{\'a}n Paredes, Carlos A.},
  title        = {qiskit-data-reuploading: a scikit-learn compatible data re-uploading classifier for {Qiskit\,2.x}},
  howpublished = {GitHub},
  year         = {2025},
  note         = {Available: \url{https://github.com/Carlosandp/qiskit-data-reuploading}}
}

@misc{tlmuavrepo2026,
  author       = {Dur{\'a}n Paredes, Carlos A. and Le{\'o}n Calder{\'o}n, Javier E. and S{\'a}nchez Perea, Nicol{\'a}s and D{\'\i}az, German Dar{\'\i}o and Segura, Camilo},
  title        = {{TLM-UAV-Quantum-Anomaly-Detection}: reproducible pipeline, notebooks, results, and figures},
  howpublished = {GitHub},
  year         = {2026},
  note         = {Available: \url{https://github.com/Carlosandp/TLM-UAV-Quantum-Anomaly-Detection}}
}

@article{deng2024ifhmnn,
  author  = {Deng, Hongli and Lu, Yu and Yang, Tao and Liu, Ziyu and Chen, Jiangchuan},
  title   = {Unmanned aerial vehicles anomaly detection model based on sensor information fusion and hybrid multimodal neural network},
  journal = {Engineering Applications of Artificial Intelligence},
  volume  = {132},
  pages   = {107961},
  year    = {2024},
  doi     = {10.1016/j.engappai.2024.107961}
}

@article{cerezo2021variational,
  author  = {Cerezo, M. and Arrasmith, Andrew and Babbush, Ryan and Benjamin, Simon C. and Endo, Suguru and Fujii, Keisuke and McClean, Jarrod R. and Mitarai, Kosuke and Yuan, Xiao and Cincio, Lukasz and Coles, Patrick J.},
  title   = {Variational quantum algorithms},
  journal = {Nature Reviews Physics},
  volume  = {3},
  number  = {9},
  pages   = {625--644},
  year    = {2021},
  doi     = {10.1038/s42254-021-00348-9}
}

@article{preskill2018nisq,
  author  = {Preskill, John},
  title   = {Quantum computing in the {NISQ} era and beyond},
  journal = {Quantum},
  volume  = {2},
  pages   = {79},
  year    = {2018},
  doi     = {10.22331/q-2018-08-06-79}
}

\end{document}